# Highly stable modular-assembled laser system for a dual-atom-interferometer gyroscope

**Chuan Sun** [1,2], **Si-Bin Lu** [1,*], **Min Jiang** [1], **Zhan-Wei Yao** [1,3], **Shao-Kang Li** [1], **Xiao-Li Chen** [1,2], **Min Ke** [1], **Jia-Hao Fu** [1,2], **Run-Bing Li** [1,3,4,†], **Jin Wang** [1,3,4], **Ming-Sheng Zhan** [1,3,4]

[1] *State Key Laboratory of Magnetic Resonance and Atomic and Molecular Physics, Innovation Academy for Precision Measurement Science and Technology, Chinese Academy of Sciences, Wuhan 430071, China*
[2] *School of Physical Sciences, University of Chinese Academy of Sciences, Beijing 100049, China*
[3] *Hefei National Laboratory, Hefei 230088, China*
[4] *Wuhan Institute of Quantum Technology, Wuhan 430206, China*
*lusibin@apm.ac.cn*
†*rbli@wipm.ac.cn*

**Abstract:**

Operating atom-interferometer gyroscopes outside a laboratory environment is challenging primarily owing to the instability of laser systems. To enhance the thermal stability of free-space laser systems, a compact laser system using fiber lasers and all-quartz-jointed optical modules was developed for a dual-atom-interferometer gyroscope. Millimeter-scale optical elements jointed on quartz plates with identical quartz supports, ensure laser power stability and facilitate component upgrades. The primary diode laser was locked to the modulation transfer spectrum of Rb atoms, and Raman lasers were phase-locked to the primary laser. Frequencies for repumping, blow-away, and detection lasers were adjusted with acousto-optic modulators. At room temperature, laser power fluctuation was under 1:1000, polarization extinction ratio exceeded 30 dB, frequency fluctuation was below 91 kHz, and phase noise reached to −100 dBc/Hz @ 1 kHz. The optical modules were tested at 5–50 °C and applied to a dual-atom-interferometer gyroscope. The fringe contrast was tested over the temperature range. The proposed system paves the way for promoting field applications of atom-interferometer sensors.

## 1. Introduction

Since their realization [1], atomic interferometers have progressively contributed to the measurement of rotation [2–7], gravity [8, 9], and gravity gradient [10–12]; facilitated tests of the equivalence principle [13, 14] and fundamental physics constants [15, 16]; and enabled the detection of mid-band frequency gravitational waves [17, 18]. Because of its high accuracy in absolute rotation measurements, an atom-interferometer gyroscope can be applied in fundamental physics [19–21], inertial navigation [22, 23], and geodesy [24, 25]. Atom-interferometer gyroscopes should be suitable for both laboratory and field applications. A laser system is essential for building an atomic interferometer gyroscope and must remain stable outside the lab. However, laser frequency and intensity are highly sensitive to environmental changes like temperature fluctuations. Therefore, developing a compact, temperature-insensitive laser system without compromising performance is crucial [26, 27].

Various studies focused on the development of highly stable laser systems have been reported to date. A previously developed integrated fiber laser system composed of fibers and fiber devices [28, 29] was characterized by stable laser intensities; however, its polarization extinction ratio (PER) was highly sensitive to temperature fluctuations. To construct a free-space integrated laser system, miniaturized metallic bases were designed and fabricated to support optical elements [30–32], achieving a reduced optical size. However, the laser intensity remained susceptible to temperature fluctuations owing to the differing coefficients of thermal expansion (CTE) of metals and glass materials. All-quartz glass laser systems, which enhance intensity and polarization stability, have been used in laser interferometers [33,34], quantum gas experiments [35], and atomic interferometers [36,37]. Using materials with identical CTEs for the base plate, supporting assemblies, and optical elements minimizes temperature influence. Nevertheless, to fulfill the stringent requirements of an atom-interferometer gyroscope for inertial navigation, a miniaturized, highly stable laser system operating over a wide temperature range is required.

In this study, we developed a compact, stable laser system using fiber lasers and all-quartz-jointed optical modules for a dual-atom-interferometer gyroscope. Optical modules were designed with heat dissipation considerations and assembled with millimeter-scale optical elements on a quartz base plate. Fiber lasers were coupled to these modules using a graded-index (GRIN) lens mounted on quartz supports. To enhance coupling efficiency, the fiber coupler was supported by inclined-plane wedges. Lasers were modulated, split, and directed to the sensor head via single-mode polarization-maintaining (PM) fibers. The primary laser was stabilized using modulation transfer spectroscopy (MTS), while other lasers were locked to the primary laser with optical phase-locked loops (OPLLs). Stability tests under varying temperatures confirmed the system's reliability, and it was successfully integrated into a dual-atom-interferometer gyroscope. Because of the proposed modular design and temperature adaptability of this laser system, it can be utilized in field applications of atom-interferometer sensors.

## 2. Design of laser system

Fig.1 illustrates the transition levels of $^{87}$Rb atoms and the laser frequencies required in a dual-atom-interferometer gyroscope. The blow-away and detection lasers are resonant with the $|F = 2\rangle$ to $|F' = 3\rangle$ transition. The cooling laser is red-detuned by 12 MHz from the $|F = 2\rangle$ to $|F' = 3\rangle$ transition. The repumping laser resonates with the $|F = 1\rangle$ to $|F' = 2\rangle$ transition. The Raman lasers have a frequency difference of 6.834 GHz and a single-photon detuning of −700 MHz from the $|F' = 2\rangle$ state. Time-division multiplexing pulse sequences manipulate the atoms because laser frequencies can be arbitrarily switched using acousto-optic modulators (AOMs).

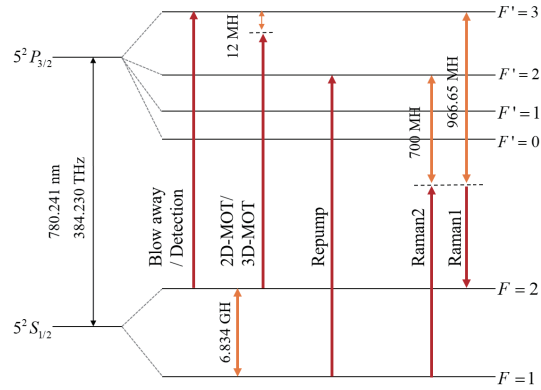

Fig. 1. Transition levels and laser frequencies for an atom-interfreometer gyroscope.

To realize the specified laser frequencies, the dual-atom-interferometer gyroscope's laser system was designed with three fiber lasers and six optical modules (Fig.2). A 1560-nm external cavity diode laser with a line width under 9.2 kHz, measured using the 50-km fiber delayed self-heterodyne method [38], was amplified by an Er-doped fiber amplifier and frequency doubled using periodically poled lithium niobate to produce a 780-nm fiber laser. These 780-nm lasers were coupled to optical modules with PM fibers. Light from the primary laser, modulated

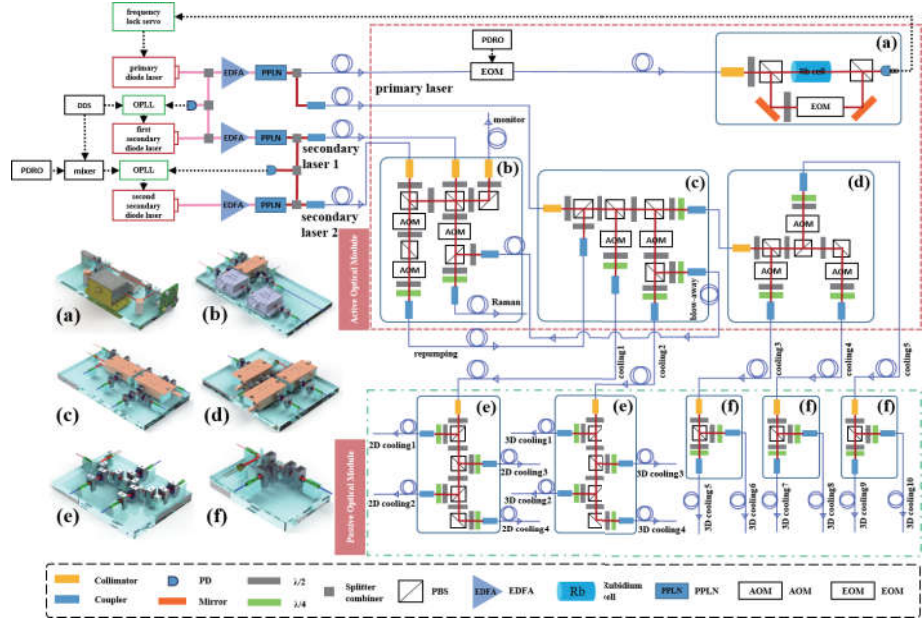

Fig. 2. Schematic of the proposed laser system. The primary diode laser is locked to the MTS and two secondary diode lasers are locked to the primary diode laser with two OPLLs. The three lasers are amplified by using Erbium-doped fiber amplifiers and frequency doubled with periodically poled lithium niobates and then coupled into the homemade optical modules with the PM fibers. The active (red dashed area) and passive (green dashed area) optical modules are separately jointed using a UV glue.

by a 1.0585-GHz fiber electro-optical modulator (EOM), was fed to module (a), wherein one beam passed through a small Rb cell, and the other beam, modulated by a 6.25-MHz EOM, counter-propagated in the cell to generate the MTS. The frequency sideband of the primary laser was locked to the $|F = 3\rangle$ to $|F' = 4\rangle$ transition spectrum of $^{85}$Rb atoms using the MTS method [39]. The first secondary diode laser was locked to the primary diode laser, and the second secondary diode laser was locked to the first secondary diode laser using the OPLL method [40]. The frequency of the primary laser was blue-detuned by 68 MHz from the $|F = 2\rangle$ to $|F' = 3\rangle$ transition, that of the first secondary laser was red-detuned by 700 MHz from the $|F = 2\rangle$ to $|F' = 2\rangle$ transition, and the frequency difference between the two secondary lasers was 6.834 GHz. The two secondary lasers were coupled to module (b). The first secondary laser and part of the second were combined using polarization beam splitters (PBSs) and half-wave plates, then split into two beams. One beam, after passing through an 80-MHz AOM, served as the Raman laser. The other monitored the beatnote signal. The remaining part of the second secondary laser passed through two 350-MHz AOMs to generate a repumping laser. The primary and repumping lasers were coupled to module (c) and combined using a PBS and two half-wave plates, and subsequently divided into four parts. One part went directly to module (d); two parts passed through an 80-MHz AOM and then to module (e) as cooling lasers. The fourth part, labeled blow away, was sent to module (b) and combined with the Raman lasers. In module (d), the combined lasers were split into three channels, modulated by three independent 80-MHz AOMs, and sent to module (f) as cooling lasers. In modules (e) and (f), the cooling light was divided into 14 beams as 2D and 3D cooling lasers, respectively. All the lasers were fed to the sensor head using PM fibers. To minimize temperature influence, active devices in the red dashed area in Fig.2 (a)–(d) and passive elements in the green dashed area in Fig.2 (e) and

(f) were separately utilized. Quartz supporting assemblies were designed in the active modules to invert and suspend the AOMs for minimizing heat conduction to the base plate.

## 3. Jointing technique for assembled modules

To achieve high integration and thermal stability in optical modules, millimeter-scale optical elements and devices were attached to centimeter-scale quartz base plates using wedged millimeter-scale quartz supports. As illustrated in Figs.2 (a)–(f), various optical components such as fiber couplers, beam collimators, beam splitters, beam combiners, optical modulators, and Rb cells are assembled in different modules using UV adhesives, shown in Fig.3 (a)–(f).

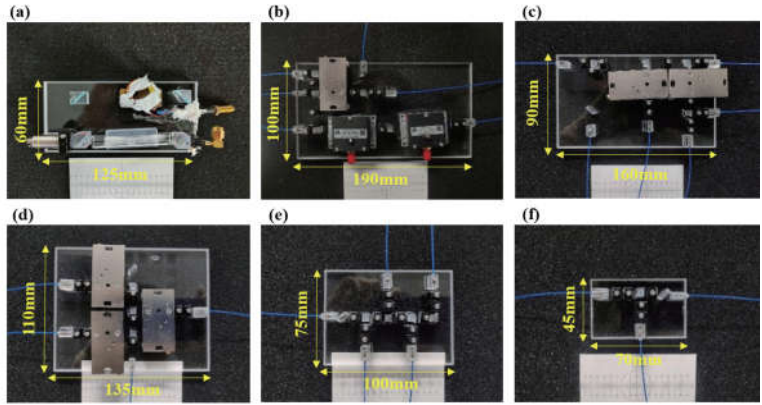

Fig. 3. Photographs of the optical modules. Millimeter-scale optical elements are jointed on quartz base plates, with active optical devices like AOM and EOM inverted and suspended using quartz supporting assemblies. The fiber coupler, fabricated by jointing a GRIN lens and a pigtailed glass ferrule into a glass sleeve, ensures beam coaxiality and fiber coupling efficiency.

Optical modules were connected using custom fiber couplers, comprising a GRIN lens and pigtailed glass ferrules aligned within a glass sleeve for beam coaxiality and efficient fiber coupling. Two quartz wedges with inclined planes supported the coupler to match height and angle, ensuring over 90% free-space-to-fiber coupling efficiency. The beam collimator was made by joining a GRIN lens with a pigtailed glass ferrule in a glass sleeve, supported by a quartz block with a groove. Rotating half-wave and quarter-wave plates before the fiber coupler ensured the laser beam's PER exceeded 30 dB after collimation. The laser beam was collimated to a diameter of 0.383 mm at $1/e^2$ to ensure that small optical elements, that is, customized PBSs with sizes of 5 mm × 5 mm × 5 mm and wave plates with a diameter of 8 mm, could be integrated into our modules. The AOM was inverted and suspended on the base plate with two quartz supports to minimize heat conduction. The first-order diffracted light efficiency exceeded 80%. In the homemade EOM, fabricated using a lithium niobate crystal supported by quartz, the crystal was jointed between two metal sheets, with the control coil fixed atop the crystal with silicone rubber. The Rb cell was held by a quartz groove and shielded with mu-metal.

To enhance the stability of the laser system, the optical modules were divided into active and passive modules, which were independently jointed. The positioning and angles of optical elements relative to the base plate were finely adjusted using a custom platform with six degrees of freedom. Optical elements were attached to the base plate with low-shrinkage, low-stress ultraviolet (UV) adhesives, allowing precise angle and distance adjustments via monitoring and prompt correction of laser coupling efficiency. This method minimized deformation during glue curing. We used four UV lights to control the jointing and preliminary curing times (set to approximately 30 min), thus minimizing stress accumulation and ensuring optimal optical performance. The optical modules were independently jointed and assembled, and a laser system with a size of 260 mm × 220 mm × 60 mm was successfully developed for atom-interferometer gyroscopes, whose size was mainly constrained by the AOMs' size and diffraction angle.

**4. Optical characterization and performance**

To assess the power stability, optical modules were tested at room temperature (25 °C). The laser was coupled to the fiber and connected to a power meter (Thorlabs, PM100D). The two output ports of the active module (c) were monitored for about 4 h at a 1 Hz sampling rate, and their relative stabilities were assessed using Allan deviations (Fig. 4 (a)). The black and orange lines represent the output laser powers (cooling 1 and 2), with stabilities better than $8.3 \times 10^{-4}$ @ 100 s. Similarly, the relative stabilities of the two output ports of the passive module (e) were evaluated based on Allan deviations (Fig. 4 (b)). In Fig. 4 (b), the black and orange lines represent the powers of the output lasers (marked as 2D cooling 1 and 3), and their stabilities are better than $4.6 \times 10^{-4}$ @ 100 s, whereas the input power stability is $1.4 \times 10^{-4}$ @ 100 s (green line in Fig.4 (a) and (b)), implying that the optical modules exhibit good stability at room temperature.

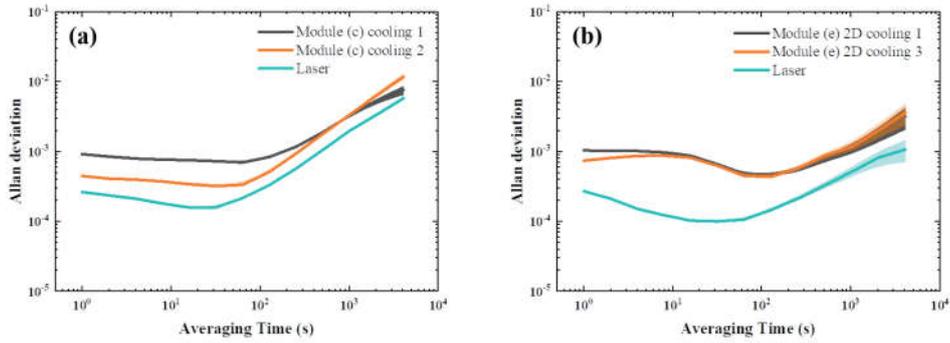

Fig. 4. Allan deviation of laser power stability at room temperature, showing relative power stability for the active (a) and passive (b) optical modules.

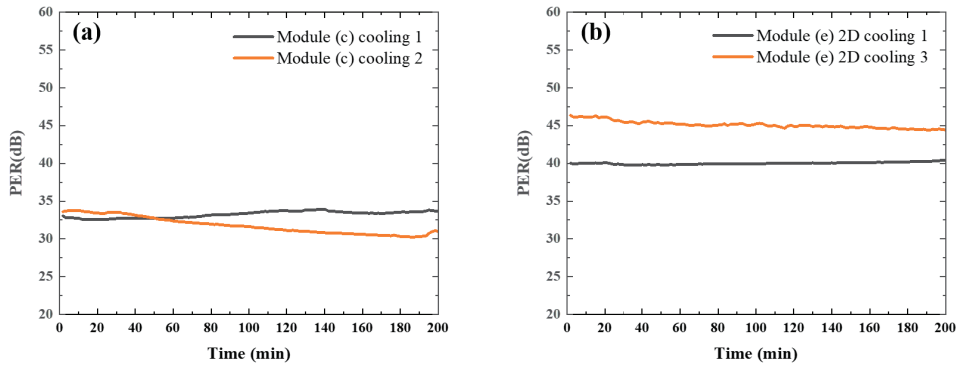

Fig. 5. Test of the PER for the active (a) and passive (b) optical modules.

The PER was measured using a polarization analyzer (Schäfter+Kirchhoff, SK010PA-NIR) at 25 °C. The output laser was coupled to the PM fiber and connected to the analyzer, with data recorded at 1 Hz. PERs for the active (c) and passive (e) modules vary over time, as shown in Fig.5 (a) and (b). Results indicate PERs of over 34.5 dB for cooling 1 (black line) and 30.2 dB for cooling 2 (orange line) with the active module (c), and over 38 dB for 2D cooling 1 (black line) and 44.5 dB for 2D cooling 3 (orange line) with the passive module (e). This demonstrates that the PER of the output remains stable over time. These results demonstrate that our optical modules satisfy the experimental requirements for atom-interferometer gyroscopes. The main limiting factors are the stability of the input laser and PER of the PM fibers.

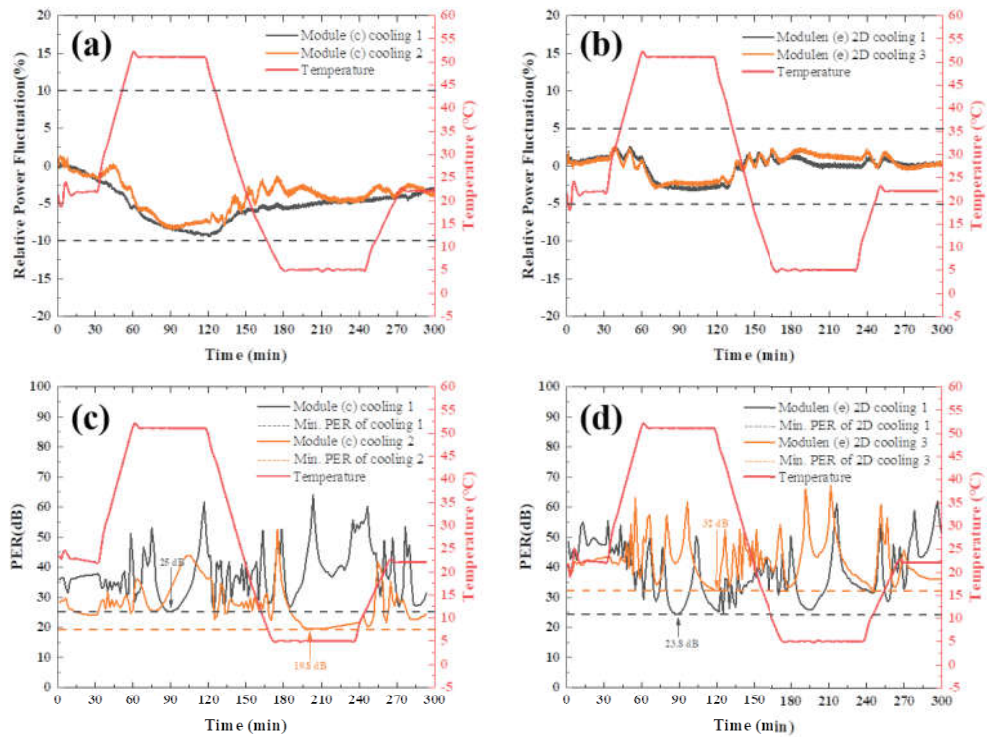

Fig. 6. Relative power stability and PER tests were conducted at varying temperatures. Stability was assessed by controlling temperature in a chamber for active (a) and passive (b) optical modules. PER was evaluated for active (c) and passive (d) optical modules.

The stability of module (c) and module (e) was tested at 5–50 °C. A temperature sensor was positioned in the optical modules, placed in the center of a temperature chamber (GAOTIAN, GT-TH-S-F900). Lasers at the input and output ports were coupled to PM fibers passing through a small hole in the chamber. The input port fiber was connected to the primary laser, while output port fibers were connected to optical power meters and polarization analyzers. The chamber door was closed and locked to prevent air exchange before starting. The data were recorded with a sampling rate of 1 Hz. The fluctuation in the relative power for the active and passive modules was tested by varying the temperature in the form of a tooth wave (red line in Fig.6 (a) and (b)). In active module (c), power decreased with rising temperature and gradually recovered as the temperature dropped, showing a peak-to-peak fluctuation of 9.4% (Fig.6(a)). Passive module (e) also exhibited a power decrease with rising temperature and an increase as the temperature fell, with a maximum fluctuation of 3.2% (Fig.6(b)). When the temperature returned to its initial value, laser power recovered by over 95%. In active modules, AOM temperatures exceeded 70 °C when chamber temperature was set to 50 °C. Power stability was affected by air heat convection despite the inverted, suspended AOMs. Fluctuations were smaller during cooling than heating, and passive modules exhibited better stability than active modules. The PER tests showed minimum values of 19.8 and 23.8 dB for the active and passive modules, respectively (Fig.6(c) and (d)), indicating superior PER in passive modules and that laser output polarization was affected by heat conduction over a wide temperature range.

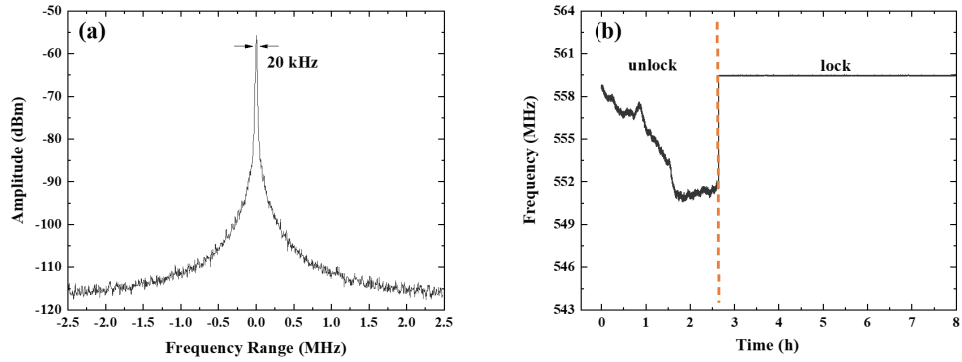

Fig. 7. Beatnote between the primary diode laser and an ultrastable laser. (a) Beatnote signal recorded using a spectrum analyzer with a 5-kHz resolution bandwidth. (b) Central frequency of the beatnote recorded for over 8 h.

The primary diode laser was locked to the MTS and tested. As shown in Fig.2, the primary laser passed through a fiber EOM and then into module (a). One beam passed through a small Rubidium cell while another, modulated by a homemade 6.25-MHz EOM, counter-propagated in the cell to generate the MTS. After locking the frequency sideband to the $|F = 3\rangle$ to $|F' = 4\rangle$ transition spectrum of $^{85}$Rb atoms, the primary diode laser was combined with an ultrastable laser. The beatnote signal was detected using an ultrafast fiber optic photodetector (Thorlabs, RXM10AF) and monitored with a spectrum analyzer (Rohde & Schwarz, FSV3000) for over 8 h. The ultrastable laser, with an average linewidth of 0.6 Hz, was locked to a 10-cm ultralow expansion cavity via the Pound–Drever–Hall method, exhibiting long-term frequency fluctuation of less than 4 kHz over 24 h. With a resolution bandwidth (RBW) of 5 kHz and a video bandwidth (VBW) of 50 kHz, the beatnote linewidth at −3 dB was less than 20 kHz (Fig.7 (a)). Given the superior performance of the ultrastable laser, the linewidth and frequency fluctuations were attributed to the primary diode laser. The primary diode laser's frequency fluctuation decreased from over 6 MHz before locking to less than 91 kHz after locking (Fig.7 (b)).

The Raman lasers was locked and evaluated. The first secondary diode laser was locked to the primary diode laser, and the second secondary diode laser was locked to the first diode laser using the OPLL method, with its performance subsequently tested. The primary diode laser was combined with the first secondary diode laser, and the beatnote signal was detected using an ultrafast fiber optic photodetector (Fig.2). This signal was amplified and connected to an offset phase-locking servo (Vescent, D2-135), with the reference signal sourced from a custom multichannel radiofrequency (RF) pulse-sequence generator [41]. The error signal was fed back to the first secondary diode laser. Once the first secondary diode laser was locked to the primary one, the beatnote signal from the servo's half-frequency division port was detected by a spectrum analyzer. With the RBW and VBW of the spectrum analyzer set to 5 and 50 kHz, respectively, the 1560-nm beatnote between the primary and first secondary diode lasers is depicted in Fig.8 (a) (black line). After locking the second secondary diode laser to the first one, the beatnote was detected using an ultrafast fiber optic photodetector, amplified, and connected to another offset phase-locking servo. The servo loop's reference signal was generated by mixing a 6.65-GHz phase-locked dielectric resonator oscillator signal with a 50-MHz direct digital synthesizer signal from a self-developed multichannel RF pulse-sequence generator. The error signal was fed back to the current of the second secondary diode laser, and the 780-nm beatnote was detected by the spectrum analyzer (Fig.8 (a) (orange line)). The phase noise from the beatnote signals is shown in Fig.8 (b). The first and second loops' phase noise values are approximately −100 (black line) and less than −90 dBc/Hz @ 1 kHz (orange line), respectively.

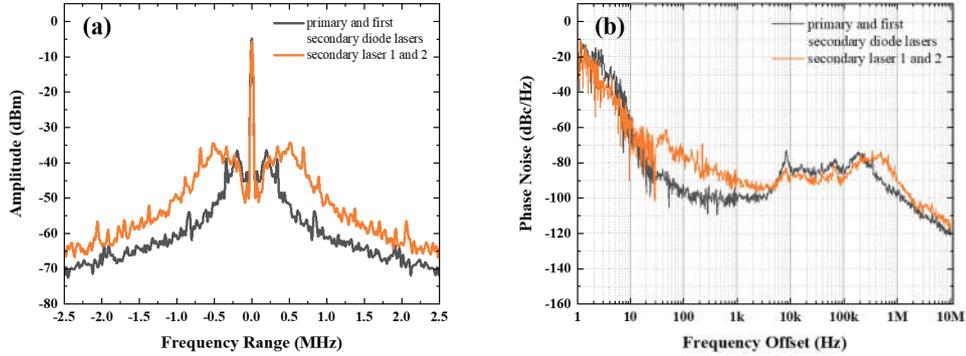

Fig. 8. Phase noise of the Raman lasers locked with the OPLL method. (a) Beatnote between the primary and first secondary diode lasers (black line) and that between two secondary lasers (orange line). (b) Phase noise derived from the beatnote signal is less than −90 dBc/Hz @ 1 kHz for the Raman lasers (orange line) and reached to −100 dBc/Hz @ 1 kHz for the primary and first secondary diode lasers (black line).

## 5. Application in a dual-atom-interferometer gyroscope

A laser system was developed for a miniaturized dual-atom-interferometer gyroscope. After testing the optical modules in high- and low-temperature experiments, they were connected, and the output lasers were directed to the sensor head using 15 PM fibers. $^{87}$Rb atoms were precooled in two 2D magneto-optical traps (MOTs) and trapped in two 3D MOTs. They were launched using moving molasses technology with horizontal and vertical velocities of 4 and 0.49 m/s, respectively. Post-polarization gradient cooling, the atoms' temperature was 8 $\mu$K. The atoms were prepared in the $|F = 2\rangle$ state owing to the delay in switching off the repump laser. The Raman and blow-away lasers selected atoms in the $|F = 2\rangle$ state and excited them to the $|F = 1, m_F = 0\rangle$ state. Counter-propagating Raman pulse sequences ($\pi/2$-$\pi$-$\pi/2$) were used to create dual-atom interferometers with a 60-ms interrogation time. Atoms in the $|F = 2\rangle$ state were detected with horizontally cooled laser beams, achieving a Mach–Zehnder type dual-atom-interferometer gyroscope [42]. To promote field applications of atomic gyroscope, the impact of temperature on an atomic interferometer was examined. A laser system and sensor head were placed on an optical table with a temperature sensor (Thorlabs, TSP01) near the laser system, and the lab temperature was regulated by a central air conditioner. The interferometer tested temperature effects, with interference fringe contrasts shown in Fig.9. In this case, the interference fringes were obtained with the co-propagating Raman pulse sequences, which avoided the influence of the wave vector of Raman lasers when the laser system was tested.

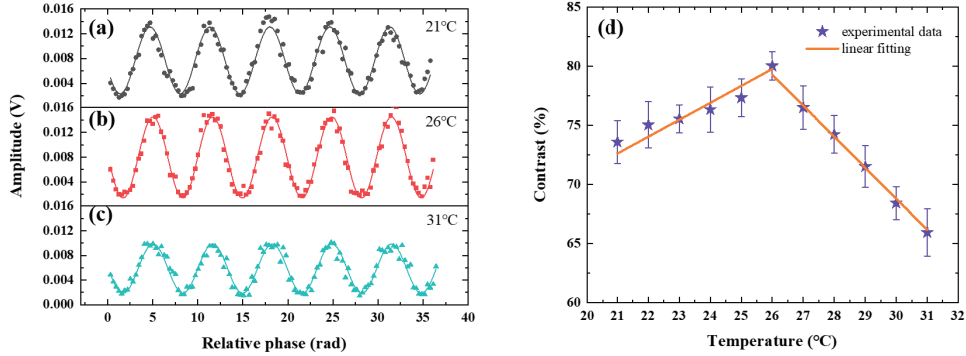

Fig. 9. Atom interference fringes at varying temperatures (a)–(c) and fringe contrast's temperature dependence. (d) Contrast diminishes with temperature changes owing to effect of the active module on the Raman lasers' power.

Optimized at 26 °C, the interferometer achieved an 80.0% contrast (Fig.9(b)), while at 21 °C, contrast was 73.5% (Fig.9(a)), and at 31 °C, it was 65.9% (Fig.9(c)). Fringe contrasts were recorded at varied temperatures (Fig.9(d)), showing maximum contrast at 26 °C. The contrast decreased linearly as the temperature decreased from 26 °C to 21 °C, with a slope of (1.4±0.1)%/°C (orange line). It also decreased when the temperature increased from 26 °C to 31 °C, with a slope of (2.8±0.1)%/°C (red line). The contrast reduction trend differed with temperature changes, aligning with the power fluctuation in the active modules, and was mainly influenced by variations in the Raman laser power, which is consistent with the test results.

## 6. Conclusion

In this study, we developed a compact, stable laser system using fiber lasers and integrated optical modules for a dual-atom-interferometer gyroscope. Millimeter-scale optical elements were mounted on a quartz plate to minimize the system's size and enhance fiber coupling efficiency. The system's modular assembly design ensures flexibility and stability. In temperature experiments, passive and active components were isolated and assembled into separate modules. At room temperature, the laser power stability and PER for the passive module were less than 1:1000 and greater than 38 dB, respectively, whereas the corresponding values for the active module were less than 1:1000 and greater than 30 dB, respectively. When the temperature was varied from 5 °C to 50 °C in the form of a tooth wave, the relative power fluctuation and PER for the passive module were less than 3.2% and greater than 25.3 dB, respectively, whereas the corresponding values for the active module were 9.4% and 19.8 dB, respectively. The primary diode laser, locked to the MTS, exhibited frequency fluctuations below 91 kHz for over 5 h. Two secondary diode lasers, locked to the primary diode laser using the OPLL method, exhibited phase noise better than −90 dBc/Hz @ 1 kHz for the Raman lasers and reached to −100 dBc/Hz @ 1 kHz for the primary and first secondary diode lasers. The laser system, applied to a dual-atom-interferometer gyroscope, showed that the fringe contrast of one interferometer, tested between 21 and 31 °C, was mainly affected by the active-module-induced power fluctuations of the Raman lasers. This modular laser system is expected to facilitate outfield application of atom-interferometer gyroscopes and can be adapted to other atom-interferometer-based sensors.

**Funding.** We acknowledge the financial support provided by the National Innovation Program for Quantum Science and Technology of China under Grant No. 2021ZD0300604; the National Natural Science Foundation of China under Grants Nos. 12104466, 12241410; the Outstanding Youth Foundation of Hubei Province of China under Grant No. 2018CFA082; the Youth Innovation Promotion Association of the Chinese Academy of Sciences under Grant No. Y201857; and Defense Industrial Technology Development Program under Grant No. JCKY2022130C012.

**Disclosures.** The authors declare no conflicts of interest.

**Data availability.** Data underlying the results presented in this paper are not publicly available at this time but may be obtained from the authors upon reasonable request.